# MODELS OF INNATE NEURAL ATTRACTORS AND THEIR APPLICATIONS FOR NEURAL INFORMATION PROCESSING

*Ksenia P. Solovyeva [1,2], Iakov M. Karandashev [1,2], Alex Zhavoronkov[3]*
*and Witali L. Dunin-Barkowski [1,2*]*

[1]Scientific Research Institute for System Analysis, Russian Academy of Sciences (RAS), Nakhimovsky prosp., 36, korp. 1, Moscow, Russian Federation, 117218
[2]Moscow Institute of Physics and Technology, Institutsky 9, Dolgoprudny, Moscow Region, Russian Federation, 141700
[3]Insilico Medicine, ETC, Johns Hopkins University, B301, 1101 33[rd] St, Baltimore, MD, 21218

## Abstract

In this work we reveal and explore a new class of attractor neural networks, based on inborn connections provided by model molecular markers, the molecular marker based attractor neural networks (MMBANN). We have explored conditions for the existence of attractor states, critical relations between their parameters and the spectrum of single neuron models, which can implement the MMBANN. Besides, we describe functional models (perceptron and SOM) which obtain significant advantages, while using MMBANN. In particular, the perceptron based on MMBANN, gets specificity gain in orders of error probabilities values, MMBANN SOM obtains real neurophysiological meaning, the number of possible grandma cells increases 1000-fold with MMBANN. Each set of markers has a metric, which is used to make connections between neurons containing the markers. The resulting neural networks have sets of attractor states, which can serve as finite grids for representation of variables in computations. These grids may show dimensions of $d = 0, 1, 2,...$ We work with static and dynamic attractor neural networks of dimensions $d = 0$ and $d = 1$. We also argue that the number of dimensions which can be represented by attractors of activities of neural networks with the number of elements $N=10^4$ does not exceed 8.

## 1. Introduction

The idea of neural systems working as a unification of many similar units ("hyper-columns") exists in neuroscience for years [1]. There are several approaches to understanding the inner machinery of elementary neural networks. One approach involves the revealing of neural connections experimentally. This is usually performed by identification of all connections in serial electron-microscopic slices of the whole brain [2]. A recent molecular engineering approach by Zador et. al. [3] appears to simplify the problem. Other approaches try to match the observations with theory. One set of ideas theorized that most neural connections are formed by associative memory processes [4-7]. The most prominent is the notion of Marr's collateral network [6], which has been later re-discovered as the Hopfield network for associative memory [8]. The network dynamics of the latter (although only in the case of symmetrical connections, which seems to be hardly possible in live neural systems) could be described by a potential energy function, which is minimized with the activity dynamics. The attractor points of Hopfield networks are isolated from each other. Continuous sets of attractor states present a completely different problem. We consider that two states $S_1$ and $S_2$ are connected, if $\rho(S_1, S_2) \leq 2$ ($\rho$ is Hamming distance). The connected set of stable states can constitute a grid of states, which can be used for representation in a brain of continuous variables. We will refer to the attractors, which can be used for the grids of the $d$-dimensional variables, as attractors with dimension $d$.

So, the set of isolated attractor points (the set of stable states in the "Hopfield network") has dimensionality $d=0$. In [9-11] the Hopfield-type neural networks, based on "continuous" sets of vectors, were considered. They can represent finite grids for one-dimensional variables [12]. The hard-wired networks (networks with innate connections) with continuous attractors are known since 1977 [13]. Recently, it has been discovered in isolated cortex slices experiments that many reverberating connections in cortex are innate [14, 15]. Also, the theoretical reasoning has been expressed that attractor neural networks can be innate and formed in ontogenesis with the help of special molecular markers [16, 17]. In this work we will present the advantages of using molecular marker based attractors for modeling basic types of neural information processing with computational experiments on attractors with dimensions $d = 0$ and $d = 1$. We will show the robustness of zero dimensional attractors to noise, we will show how can be visualized all states of preformed linear ring attractors ($d = 1$), and we will then analyze how the dimensionality affects the learning of an attractor network and present an extension of Kohonen's SOM. All the learning experiments will be done with McCulloch-Pitts Neurons. Some of the experiments with the activity dynamics in neural networks with innate connections will be extended to Leaky Integrate and Fire Neurons to show that the presented concept is not limited to one neuron model.

## 2. Attractor Neural Networks

There are at least three general mechanisms for making attractor neural networks. The first is the self-obvious method [13]. Here, the neurons are considered to be located in physical space and the connections are established in direct relation to the distance between neurons.

The second mechanism uses Hebb modifiable synapses. For $d = 0$, it was proposed in [6, 8]. For $d = 1$, it was studied in [11]. To make the neurons of the network properly interconnected, they should be exposed to the signals from the external world for a certain period of time. This is provided by extensive scanning of the environment by the animal hosting the neural network [20, 24]. In this "conditioning" process, the neurons, which get similar information from the external world, are often excited simultaneously. Due to Hebb-type learning rules, they become connected. Thus, in associative neural networks, the firing of each neuron is connected to the specific input information by the inborn connections.

Here, we propose and explore the third mechanism. The idea has been discussed earlier in [18]. Our approach makes use of the following considerations. There is undisputed data showing that many inter-neuronal connections are inborn [14, 15]. The inborn attractor networks are obtained with the help of connection rules, enabling neural networks to have attractors with the desired properties.

In contrast to the attractor networks, based on Hebb synapses, the attractor neural networks with inborn connections inside them must tune their external connections to endow the neuron firing with a certain sense. We consider concrete examples of such processes later in the paper. In this paper, we deal with the networks with inborn connections inside the network and restrict analysis to the attractor dimensions $d = 0$ and $d = 1$.

## 2.1 Molecular Marker-based Attractors, $d=0$

Molecular markers can be used to get the matrix for the neural network with $M$ isolated ($d=0$) attractor points. There are $M \times L$ markers belonging to $M$ classes with $L$ elements in each class. The distance between markers is 0 when markers belong to the same class, and non-zero, say $2L$, when markers belong to different classes. The markers are distributed randomly between the neurons so that each neuron gets $q = (M \times L) / N$ markers, which all belong to different classes. For simplicity, we consider only cases when $q$ is an integer while generalization to non-integer values of $q$ is not difficult. Then, the neurons $i$ and $j$ are connected with excitatory connections only if the neurons contain markers of the same class. Contrary to the method of the learned connections, this method does not specify which concrete states belong to the attractor in advance. The set of attractor states depends on the results of the random distribution of markers between neurons. This connection rule provides mutually excitatory connections between neurons that have the same type of markers. This means that neurons with same type of markers might persistently excite each other. The state of the network when these $L$ neurons are excited and the rest neurons of the network are silent, presents an element of the set of the neural network attractor states. This statement holds for each of the $M$ types of molecular markers. Therefore, the attractor for the neural network in which neural interconnections are made with molecular markers consists of $M$ states, $S_m$, ($m = \overline{1, M}$) of activity of the network of $N$ neurons. This is true (with the probability close to 1) , while $M$ does not exceed certain value, which depends on $N$ and $L$. When $L$ is small compared to $N$, the distance between any two states of the attractor is close to the value $2L$. The interconnection matrix consists of 1 and 0. It is symmetric and has all zeros at the diagonal. The Hebb-Hopfield method [11] and the method of molecular markers result in the similar matrices. The only difference is in our knowledge of attractor states. In the first case, we initially select states of network activity which are to be the attractor states. In the second case, we randomly distribute markers between neurons while not knowing which neurons will be active in that or any another attractor state of the network.

More details on MMBAN with $d$=0 are given in Supplementary Material 5.

## 2.2 Bump Attractors $d$=1

The first neural network with a continuous one-dimensional attractor has been was discovered by Sun-Ichi Amari in 1977 [13]. He considered a set of neurons located along a line with excitatory local and inhibitory more distant connections ("Mexican hat") [13]. In this network, there are stable states of activity in which neurons of a local group are active and the rest neurons are inhibited. This type of attractor is known as a "bump" attractor, as active vs. inactive neurons in the attractor state present a "bump" on a line of neurons. These types of attractors exist in neural networks of many types of neural models. For convenience, we give the general definition of this phenomenon below.

*Definition of bump attractor.* Consider a network of $N$ MCP (Supplementary Material 1) neurons which are connected to each other. For handling the neural network, it is substantial that each of the neurons has an individual identity. Without loss of generality we can consider that these identities are order numbers, from 1 through $N$, each of which is permanently attached to a concrete neuron. This attachment provides neurons with their "individual names". Ordering the neurons according to these numbers yields the basic order of neurons. Sometimes, it is convenient to use the altered order numbers of the same neurons. The particular alternate enumeration of neurons can be presented as the set $\{a(1),...,a(N)\}$, which is a permutation of $\{1,...,N\}$. The $a(i)$ can be also considered as a vector-function, i.e. a mapping of the set $\{1,...,N\}$ into itself. Obviously, for $N$ neurons there could be exactly $N!$ enumerations. The state of the network, in which $L$ neurons are permanently active and $(N-L)$ neurons are permanently silent we name the attractor state of the network. Note that for any attractor state $s = (s(1),...,s(N)); \ s(i) \in \{0,1\}$, with $L$ excited elements ($s$=1) of any network, there exist such an enumeration $\{a(1),...,a(N)\}$ that $s(a^{-1}(L+1)) = ... = s(a^{-1}(L+L)) = 1$, where $a^{-1}(i)$ is the reverse function to $a(i)$. In other words, for any attractor state of the neural network, there exist such an enumeration of neurons with which all excited neurons are numerated sequentially, starting with order number $(L+1)$. This fact is indeed trivial, but we need it for further formulation. Now, we say that all attractor states of the network constitute the bump attractor, if for each attractor states of the network there exists such enumeration, $a_s$, that for the states $s_i$; $(i = 0, ..., 2L)$ keeps correct that $s_i(a_s(1+i)) = s_i(a_s(2+i)) = ... = s_i(a_s(L+i)) = 1$ and these $(2L+1)$ states are attractor states of the network. This rather complicated definition can be clarified by the graphics of Fig. 1. The case of continuously-valued LIF (Leaky Integrate and Fire) neurons is also shown here.

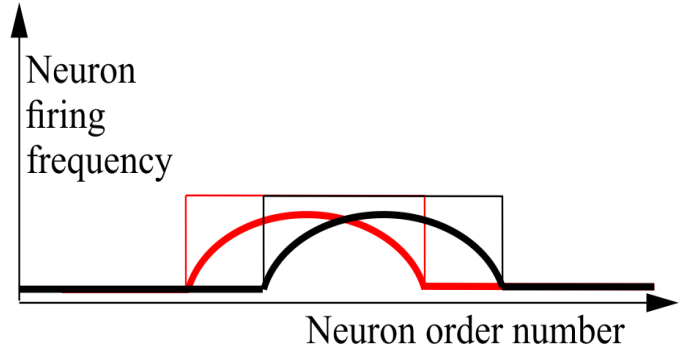

**Fig. 1. To the definition of one-dimensional bump attractor**
Red and black are adjacent stable states of bump attractor networks for proper enumeration of neurons in the network. Thin line - MCP neuron model, thick line - continuous-valued neuron model.

In Fig. 1, states of individual neurons are plotted vs. their order number. For the neural network with bump attractors, for each attractor state there exists an enumeration with which there are at least $2b+1$ attractor states $b \sim L/2$, each of which starts with excited neuron at order number $i = L - q + 1$ and continues with excited neurons through the order number $i = 2L - q$, $(q = 1, ..., b)$. Thus, each of these $2b+1$ attractor states in this enumeration looks like a "bump" of $L$ successive active neurons, while the rest of neurons are silent. This definition is practically trivial for the network, where all neurons constitute a ring of the length $N$ with excitatory connections between adjacent neurons. In these rings, the number of attractor states (the "length" of the ring of attractor states) coincides with the number of neurons in the network $N$. Later in this paper we will demonstrate that "the rings" of attractor states can be of the length $kN$, where $1 < k < K(N)$, with $k(N)$ linear growing with $N$ with constant $L$. In other words, the fact that a neural network has a bump attractor means that the set of its attractor states can be presented as locally linear in vicinity of each of its attractor states. To observe this presentation, the appropriate enumeration of neurons should be selected. In general case, the necessary enumeration depends on the concrete state in vicinity of which we make observable the local linear structure of the attractor. However, for some types of bump attractors, there do exist enumerations which yield linear representation for substantial part $(1/R)$-th of all attractor states; $R = 1, 2, 3, ...$ depending on the case) [10, 11]. In this paper, we do not consider bump attractors with dimensionality $d > 1$. However, the generalization of the bump attractor definitions to these cases can be more or less straightforward.

The bumps can be either stationary for stationary attractor states [11], or propagating over the line of neurons in case of dynamic attractors [10, 11]. The formal definition for dynamic bump attractors can be simply obtained from definition of a static bump attractor. In the case of a dynamic attractor, the velocity of propagation of the bump over the line of neurons can vary, depending on the excitatory or inhibitory background [9, 11, 21]. The number of attractor states in bump attractors can exceed the number of neurons [8]. The properties of learned bump attractors, which emerge in process of network connection forming after learning in the network of activity patterns, are described elsewhere [9, 11]. Next section deals with inborn bump attractors.

### 2.2.1 Pre-formed Attractors, $d$=1

An inborn mechanism to obtain neural networks with $d=1$ bump attractors has been proposed [16, 17]. Computational tests of these mechanisms have been first tried in [28]. This network consists of $N$ interconnected binary neurons. Its state is characterized by N-dimensional vector of "0" and "1". The interconnection matrix $\mathbf{T}$ ($N \times N$) is formed with the help of model molecular markers, as is described below. For network dynamics computing we use the asynchronous random dynamics, which is defined in Supplementary Material 1. The general idea of molecular markers in this section and in section #2.1 is similar , but the details differ, and help in building a neural network with attractors with $d$=1, instead of $d$=0.

Here, the model molecular markers $\mu_h$; $h = 1,...,M$ are considered to make a ring, so that the marker next to the marker $\mu_M$ is $\mu_1$. For the distance between markers $\mu_i$ and $\mu_j$, we take $D_{ij} = \min\{|i-j|, (N-|i-j|)\}$. We assume $M=kN$, with integer $k$. Then $M$ markers are distributed between $N$ neurons randomly, providing each neuron an even number of markers, $k$. Besides, for markers, which fall into one neuron, we demand that the distance between the markers exceeds a fixed value, $\Delta$. This process is schematized in Fig. 2. Here, the resemblance between the markers is denoted by the colors of the "molecular markers".

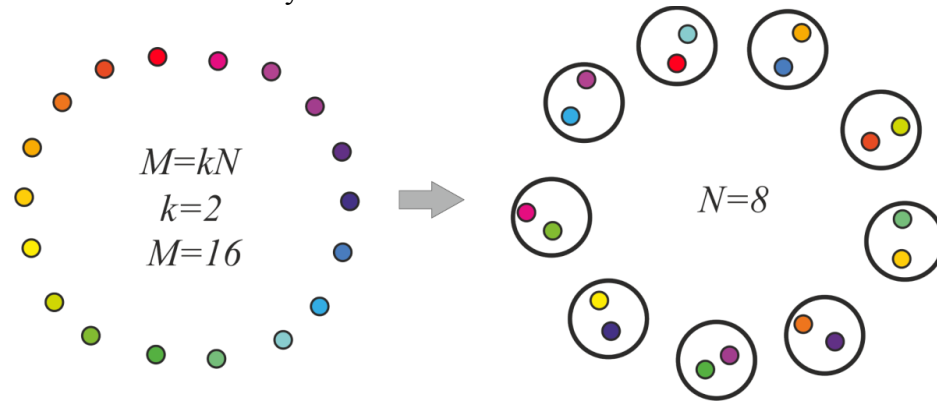

**Fig. 2. Model molecular markers (left) and their random distribution between neurons (right)**

The distribution of markers in neurons is performed by placing one marker at a time into one neuron with the help of random number generation and necessary checks. This process either yields a valid distribution of markers in a limited time, or it does not. The larger the values of $N$ and the smaller $M$ and $\Delta$ are, the sooner the distribution process is completed. After completion, those neurons, which have markers with distance less than δ, establish excitatory connections with each other with connection weight value of +1. The rest neurons are also connected with inhibitory connections of the weight -σ. The neurons are not connected to themselves. In the network, which has been formed as described, there are stationary attractor states. Several methods can be used to characterize the structure of the set of attractor states. The first one uses statistics of the states, into which the network gets after transients ("relaxes") from randomly chosen states. It is well known that the network with symmetric connections gets to attractor states in $\sim \lg(N)$ time steps [8]. The second method makes use of "artificial" dynamics introduced into the neural network. In this case, the neural threshold is made dependent on the integral value of the recent activity of the neurons [11]. Due to the threshold accommodation, the activity of neurons tends to shift from the current to the adjacent state. This method enables the activity of the network to 'slide' over any connected chain of states which exists in the neural network. In case of the closed ('circular') chains, the activity can circulate over the circles for an indefinitely long time. It should be stated here that in this paper we are dealing with sets of markers and sets of states, which compose rings. However, the ring structure is used here only for convenience of avoiding setting of boundary conditions in the beginning and the end of the chains of states or markers.

### 2.2.2 Extension to Kohonen's SOM

As has been stated in the beginning of Section 2, the neural networks, which have inborn inter-neuronal connections, must tune their external connections to endow the neuron firing with certain sense. Our approach is based on a natural extension of Kohonen's SOM approach [31] to neural networks, which have one-dimensional attractor state.

The model consists of $R$-dimensional input space and a neural network of $N$ neurons. For convenience, we consider that $R$ input "receptors" read out the cyclic input variable, $\varphi$, which might be (for example) the head direction angle $0 \le \phi \le 1$; as $\varphi$ is cyclic, $\phi = 1$ represents the same direction as $\phi = 0$. At a given value of angle $\varphi$, receptors, which are broadly tuned to this angle, are excited. The tuning curve of each receptor is bell-shaped with a definite width. The time is discrete. Values of the input variable in successive moments of time are independent and randomly selected.

In the naive system, all receptors are connected to all neurons of the targeted neural network, and connection weights are randomly chosen. The neural network consists of $N$ binary neurons with recurrent symmetric connections. In computational experiments, two types of recurrent networks are used. In the first series, we use the network described in section 2.2.1 with $k=1$. We will refer to such a type of network as a network with a full ring attractor. In the second series, we use the network which has two independent full ring attractors [12, 32]. Such networks have two independent (random in relation to each other) ring attractors of length $N$ each.

The learning process includes the following steps:

    1. The value of the input variable is selected;

    2. The receptors are excited according to their tuning curve, yielding the $R$-dimensional input vector for the neural network; [no activity in the network]

    3. The neurons are activated due to input signals from the receptors via the connection matrix;

    4. In the series of iterations, the states of the neurons of the network are updated, starting with the state, obtained at the previous stage, in accordance to the neural network equations for 20 units of time. [no working receptor connections, network activity turns to stable in 3-6 time steps ]

    5. The connections from receptors to the neurons are modified and stages 1-5 are repeated.

The rule for modification of connections is following:

$$\mathbf{W}_{i:} = \mathbf{W}_{i:} + \eta \cdot ((\mathbf{X}\mathbf{V}^T)_{i:} - \mathbf{W}_{i:})$$

This expression means that $i$-th line of the $R \times N$ matrix $\mathbf{W}$ is slightly turned (as scaled with a small parameter $\eta$) in direction of the input vector, $\mathbf{X}$, which has elicited the current state of the network activity, $\mathbf{V}$.

The modification of connections continues until it could be obvious that the matrix **W** yields continuous mapping of the variable, which is sampled by the receptors into the attractor states of the neural network. This conclusion is made based on a visual observation of the matrix **W**.

## 3. Experiments and Results

### 3.1 Robustness to Noise, $d = 0$

In this experiment we observe the effects of noise when input vectors are randomly chosen. First, we construct the attractor neural network with $M$ "isolated" attractor points, as described in section 2.1. Then, we select and fix $M$ random vectors in $R$-dimensional space. All coordinates of vectors are selected randomly from the interval [-1, 1]. Initial values of connections from input fibers to the representing neural network are random. One by one, we feed all selected random vectors to the input fibers and memorize the states of the neural network to which the network states converge with inputs from each input vector. At this stage, the input vectors are selected with the following considerations . If the newly fed preselected vector imposes convergence of the neural network into the state, which has been already memorized for a previously fed vector, we select a new random vector with which the system converges to the previously non-'occupied' attractor state. When the selection procedure is complete, each of the attractor state has a corresponding to it random vector in the input space. Afterwards, we feed the same input vectors with noise added to the network. Fig. 3 A shows results of tests of the network with the selected $M$ vectors and with same vectors with added noise. In this case, to each selected vector $V_i$ we added noise, i.e. the $R$-dimensional random vector $\xi_i$, which coordinates are random in the interval $[-\eta, \eta]$, where $\eta \in [0,1]$ is the noise amplitude. Two types of the network are compared. One of them is the network with $M$ ("isolated") attractor points. The other is the network of non-connected neurons. From those 'non-connected' neurons for the given input vector, $L$ neurons, which receive maximum excitation, are set excited.

For learning of connected and not-connected networks, the modified Rosenblatt's perceptron rule was applied. Perceptron is a device, which for each of the selected vectors $X_t$ calculates values $O_t$:

$$O_t = \sum_{i=1}^{R} w_i X_{ti}$$

where $w_i$ are the tunable real-number-valued parameters of the perceptron. The teacher compares the sign of $O_t$ with $y_t$. If they coincide, the vector of $w_i$ remains the same. If they differ, the vector of $w_i$ obtains the new value (Rosenblatt's rule):

$$w = w + X_t y_t$$

The perceptron learning is fast. It is known that if this iterative learning process converges to fixed values of $w_i$ the final state is attained after only a few iteration steps [26]. We use a modified perceptron learning rule as described below. In each cycle of work of the system there are two phases of functioning. In the first phase, the input vector $X_t$ acts on all neurons of the network, imposing a state of excitation on some of them. $L$ neurons, which get maximum excitation from the input, are left excited at this phase. The second phase of the work of the system is a relaxation of the system from its initial state, to one of the attractor states of the network. Afterwards, the connection weights of the neurons with the external fibers are modified. For those neurons in which the initial states coincide with the final state, and the final initial states coincide with the final no actions are undertaken. For neurons, which initial state was 0 and final state 1, the $X_t$ vector is added to its external connections vector. For neurons, which initial state was 1 and final state is 0, the $X_t$ vector is subtracted from its external connections vector. This procedure is repeated for all input vectors $X_t$ until matrix $W$ keeps changing. In our computations, the number of iterations in all cases was less than 50.

As can be seen from Fig. 3B, learning dramatically changed the noise dependence of the attractor neural network. Up to very large noise, neurons keep discharging with the same pattern. The behavior of uncoupled neural network does not substantially depend on learning (note huge scale difference in abscissa of Fig. 3 A and 3 B).

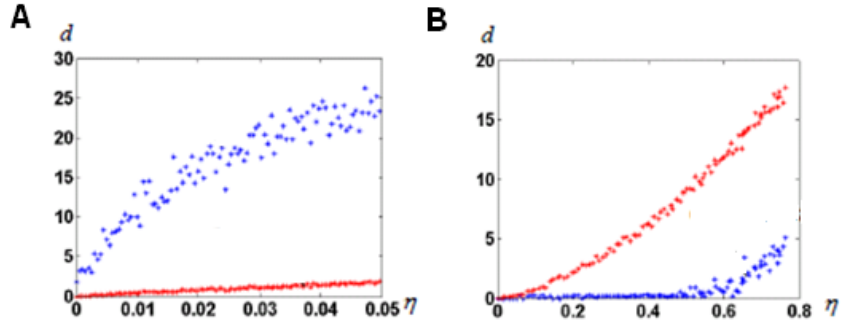

**Fig. 3. Noise dependence of "output error" of neural network responses**
A. Before learning; B. After learning. Blue dots – attractor neural network, red dots – not connected neurons. Note the huge difference of abscissa scales between A and B. $R$=100; $N$=300; $M$=100; $L$=20.

We explored this phenomenon in a wide range of parameters. In all cases, qualitatively, the behavior was the same. Neural networks with attractors demonstrate high tolerance to noise (up to 50%), while uncoupled networks are not resistant to noise. The revealed phenomenon, although transparent in its mechanisms, clearly demonstrates salient advantages of attractor neural networks with $d = 0$.

### 3.2 Preformed Attractors, $d$=1. Visualization of Ring Attractors

Fig. 4 gives an example of activity in a neural network of 300 neurons, whose connections were made with a help of a ring of 900 markers as described in section 2.2.1. We introduced "artificial" dynamics in the network, as described in 2.2.1. The plot color codes the distance (in configurational space) between the current state of the network and the states in the past (above the mid-line) and the future (below the mid-line). This method of visualization of multidimensional processes, $L$-plot, was described in [29].

The horizontal lines above and below the mid-line show that the activity in the network is cyclic, with the period of $T_{net}$ = 1225 time steps. The periodicity of the neural network activity means that there is a closed chain of attractor states in the network, all of which are attended by the system in cyclic dynamics. The period of the cyclic activity is determined by two factors: (a) the number of states attended by the system, and (b) the rate of threshold accommodation. The plots of Fig. 5 are obtained by averaging of the plot of Fig. 4 over the Ox axis. Top and bottom graphics show the same data with different time scales. At the top plot the main feature is presence of the central "negative impulse" and its two symmetric replicas "in past and future". They simply reflect the (almost) periodic processes in the neural network. At the bottom of Fig. 5, one can see that in the time segment $-10 \le \Delta t \le 10$ the derivative, $dr/dt$ is in fact constant, changing sign at $t = 0$. The important fact is that the product of $T_{net} \cdot (dr/dt)$ in this case is $1225 \cdot 1.47 = 1800.75$, which practically coincides with the value $2M = 2 \cdot 3N$, the doubled number of markers, which were used for forming connections in the network in this case. In other words, this neural network has $M$ attractor states that are connected into the ring chain of states with a minimal distance $D_{min} = 2$ between adjacent states. All of them can be visited sequentially if neurons have the property of threshold accommodation. The form of the curves at Fig. 5 can be qualitatively explained. First, we explain the initial linear growth of the distance from the given state to the subsequent attractor states as a function of time. The growth is due to the fact that all attractor states have the same value of $L$ (the number of "ones" in state vector), and the activity sequentially runs over all attractor states. The plot in fact means that activity propagates in the ring of attractor states with the same local properties, as activity propagates over a line of neurons.

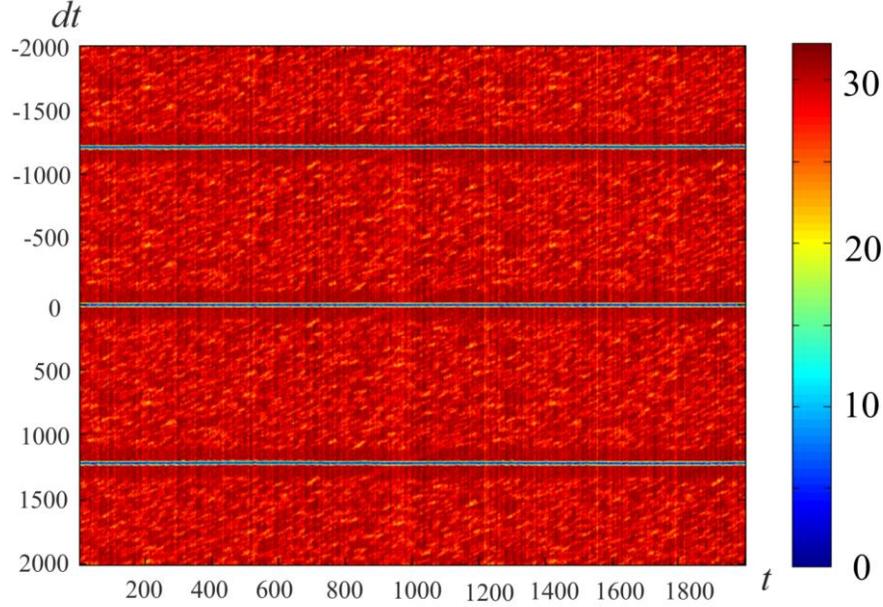

***Fig. 4. Activity visualization (L-plot [26]) in the network with connections, based on the ring of markers***

$N = 300$, $M = 900$, $\Delta = 80$, $\delta = 12$, $\sigma = 3$, mean value of $L$ over the observation period, $\overline{L} = 15$.

As this linear increase of $r$ continues almost until $r = 2L$, a slight distortion of linearity can be seen close to the value $r = 2L$. Afterwards, the distance remains equal to $2L$ (which would be exactly the same as in case of activity propagation in a linear chain of neurons). However, the distance drops to another level, $D$, when the number of the states, passed by the network activity, from the reference state approaches $\Delta$. In computational experiment, we have $D = 29.06$. Theoretical estimates yield the following expression for $D$ (Supplementary Material 3):

$$D \approx 2L(1 - \frac{(r-1)L}{M}) \qquad (4)$$

for $L = 15$, $k = 3$, $M = 90$ (4) gives $D \approx 29.06$, in good accordance with the experiment. Another way of activity visualization in this type of neural network is presented in Fig. 6. Here, abscissa gives the order number of the specially selected M neuron states. Each of them includes $L$ excited neurons, which contain markers with order numbers $i$, $i+1$, $i+2$,..., $i+L-1$, with $i = 1,...,M$. Ordinates give discrete time, which increases from top to bottom. The color of the point $(i,t)$ at this plot gives the distance between the current state $s(t)$ of the neural network and the state $s(i)$ as defined above. Fig. 6 shows that, in fact, the network states "slide" over the set of the selected states. The plot of the type of Fig. 6 is characteristic for the computational experiments with sufficiently small value of $k = M / N$.

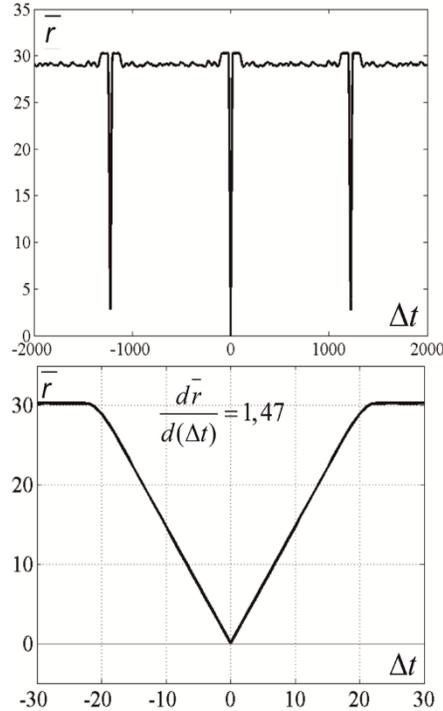

**Fig. 5. Average distance between the current state and past and future states**
Top and bottom differ in time scale. $N = 300$, $M = 900$, $T_{net} = 1225$.

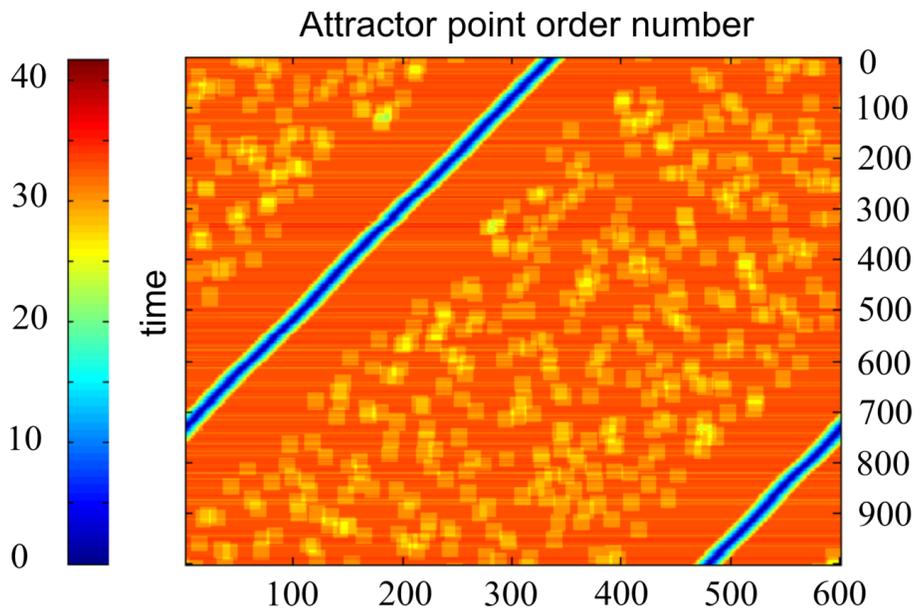

**Fig 6. Time plot of distances between the current activity state and candidate activity states, enumerated by the marker order number**
Abscissa – network state order number; ordinate – time;
$N = 300$, $M = 600$, $D = 80$, $\delta = 12$, $\sigma = 3$.

### 3.3 Activity of Neural Networks for different values of $k$

We studied effect of $k$ on network activity. With larger $k$, the neural network activity display can show the pattern given in Fig. 7. It can be seen that for $N = 300$, there exists a critical value $k = k_c$, such that for $M < k_c \cdot N$ the network activity follows the pattern displayed at Fig. 6. Screening of the parameter values in computational experiments yields the dependence of $k_c$ on $N$ (Fig. 8). It is practically linear. An analysis gives the following expression (Supplementary Material 4):

$$k_c(N, \Delta, \delta, L) = \frac{N}{2\delta \sqrt[L]{N}} \qquad (5)$$

Correction factor $\sqrt[L]{N}$ in the denominator is between 1.0 and 2.0 for $L \geq 20$ and $N \leq 10^5$. The correspondence between theory and computational experiment is fair.

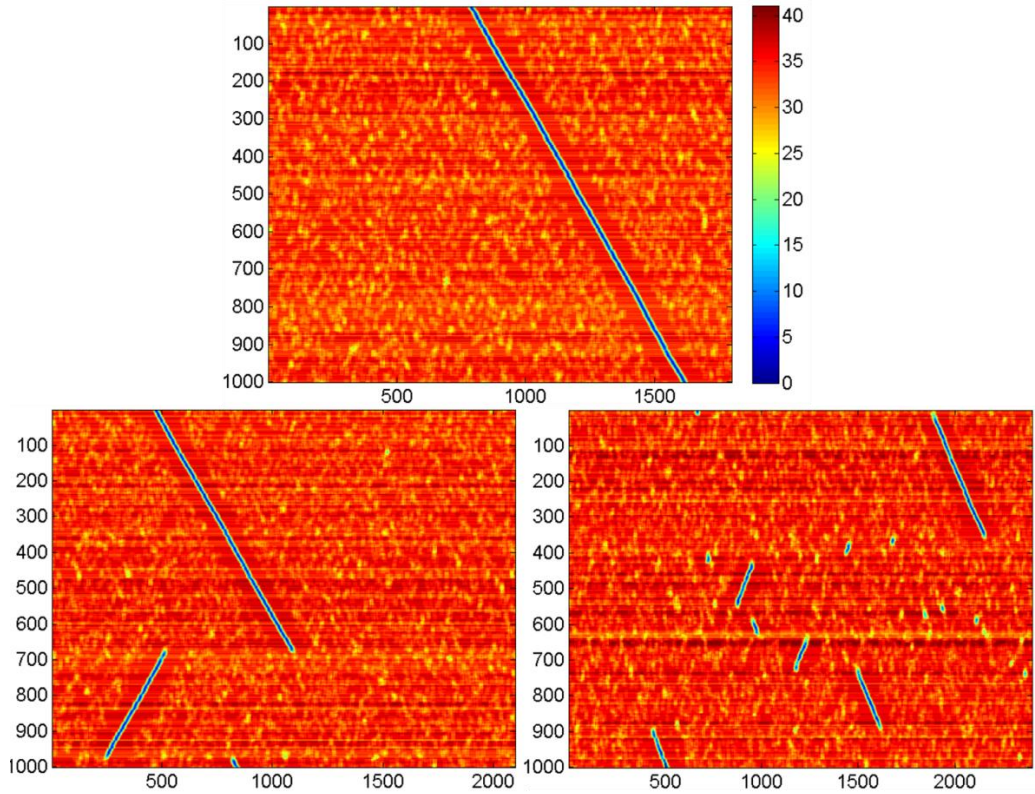

**Fig. 7. Activity display for the neural networks formed with different values of $k$**
Notations are the same as in Fig. 6. $N = 300$, $D = 80$, $\delta = 12$, $\sigma = 3$. Top: $k = 6$, in this case the activity run in cycles over all $k \cdot N$ attractor states (not shown); bottom left: $k = 7$; bottom right: $k = 8$.

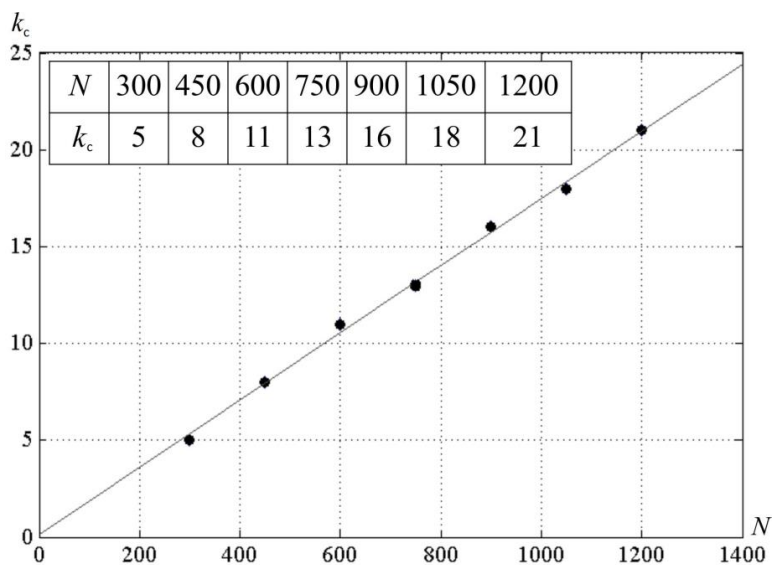

| $N$ | 300 | 450 | 600 | 750 | 900 | 1050 | 1200 |
|-----|-----|-----|-----|-----|-----|------|------|
| $k_c$ | 5 | 8 | 11 | 13 | 16 | 18 | 21 |

**Fig. 8. The dependence of $k_c$ on the number of neurons in the neural network $N$**
$D = 80$, $\delta = 12$, $\sigma = 3$; a line is a least square match of the computational experiment data.

### 3.4 Learning in Static Bump Attractors

Fig. 9 A and B show connection matrices before beginning and after completing the learning process. In the initial state, the matrix presents a random mosaic. It should be noted that the order numbers of receptors are given according to the variable values to which the receptors are tuned. Fig. 9 B shows the fact that learning **W** implements the continuous mapping of the sampled variable into attractor states of the neural network. Fig. 9 C and 9 D show the results of testing the learned system. The test consisted of sequential presentation to the system of $\varphi$ values in the range [0, 1] with 0.001 steps. With each $\varphi$ value, the neural network relaxed for 20 steps of time to an attractor state. In Fig. 9 C the final state of the network for each $\varphi$ value is represented by a line with light blue points for active neurons of the final state and dark blue points for the silent neurons. It can be seen that the sequential test of a 1000 values of $\varphi$ in the learned system yields activation of the sequential attractor states of the neural network. Fig. 9 D shows this result rather differently. Here, the inner circle of the figure represents the $\varphi$ values. The outer circle represents the order number of the neural network states to which the activity of the neural network converges when it is activated with the concrete $\varphi$ value. The latter are connected to the former with thin lines. This form of the graphic display of the representation of input variables in the neural network shows more detail than the method in Fig. 9 C. In particular, it can be seen that there are some deflections from the linear relations between the $\varphi$ values and the attractor states. Fig. 10 shows results of learning for the case when the neural network has two independent full ring attractors. The inter-neuronal connections are formed with help of two sets of molecular markers. Each set has $N$ elements with circular topology. Both sets are distributed randomly between the neurons so that each neuron gets one marker from each set. The excitatory connections are made between neurons which have markers (of either type) with distances less than $\delta$. Fig. 10 A and B give two views of the neural network interconnection matrix. These are obtained with two different enumerations of the neural network neurons. The two enumerations correspond to two different sets of markers. The views seem to be identical, but they are in fact completely different in their fine details. Fig. 10 C and D give two looks at the matrix **W** between the receptors and the neural network after learning. For the neurons, the enumerations of Fig. 10 A and B are used for Fig. 10 C and D. The first view shows that the learning has provided mapping of the variable $\varphi$ onto one of the ring attractors of the neural network. Which of the two attractors finally "accepts mapping" depends on the random initial conditions.

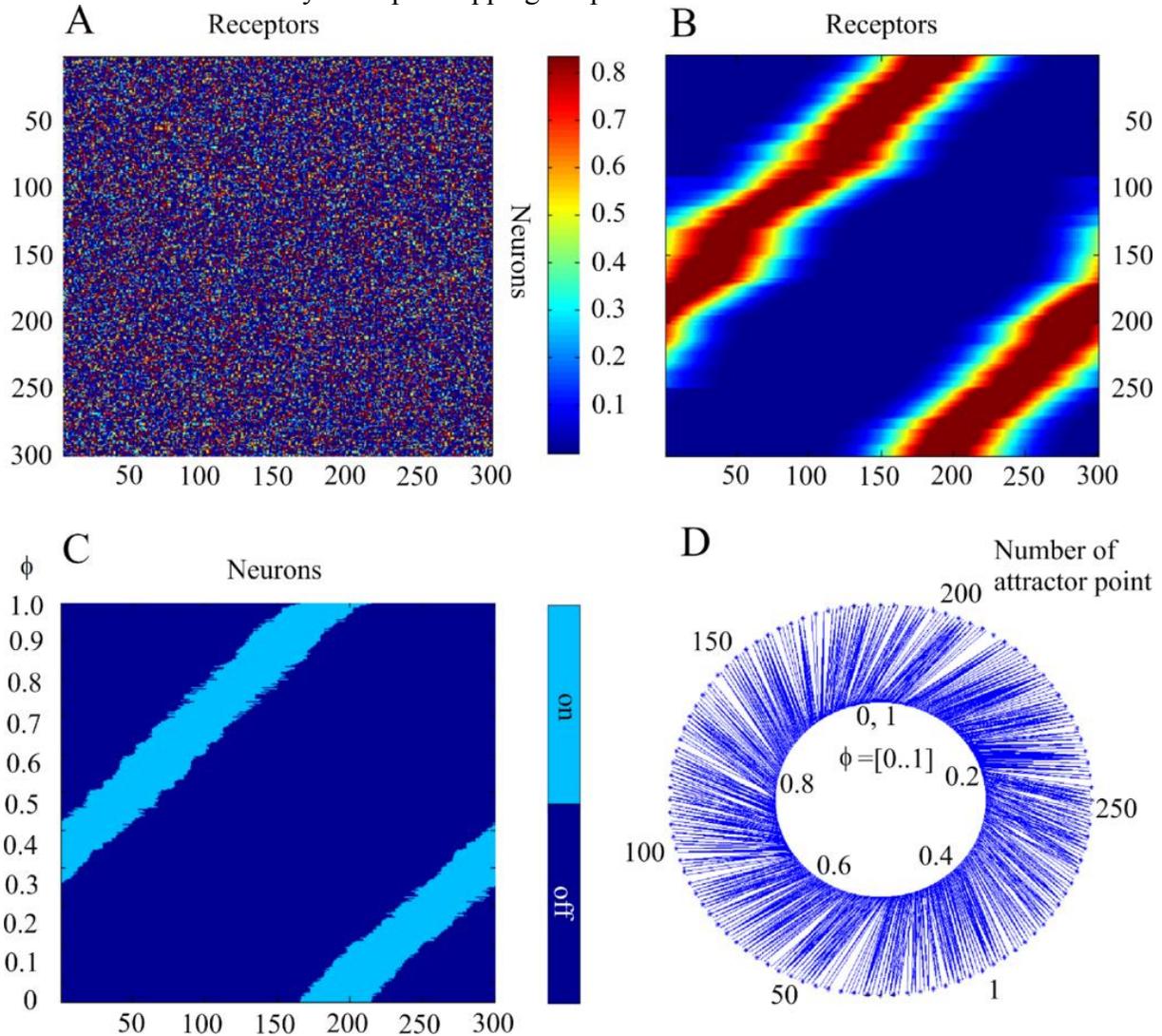

**Fig. 9. The self-organization of mapping from receptors to the attractor network**
A. Initial state of the matrix of the connections between the receptors and neurons of the attractor network. B. The same matrix after completion of self-organization. C. States of the network to which the activity of the network converges when the parameter of input signal takes 1000 sequential values in interval [0, 1.0]; abscissa – order # of neurons, ordinate – values of the parameter; light blue are excited neurons. D. Schematic mapping of parameters of the input signal to the states of the network. Inner circle – input signal parameter, outer circle – the attractor state order #.

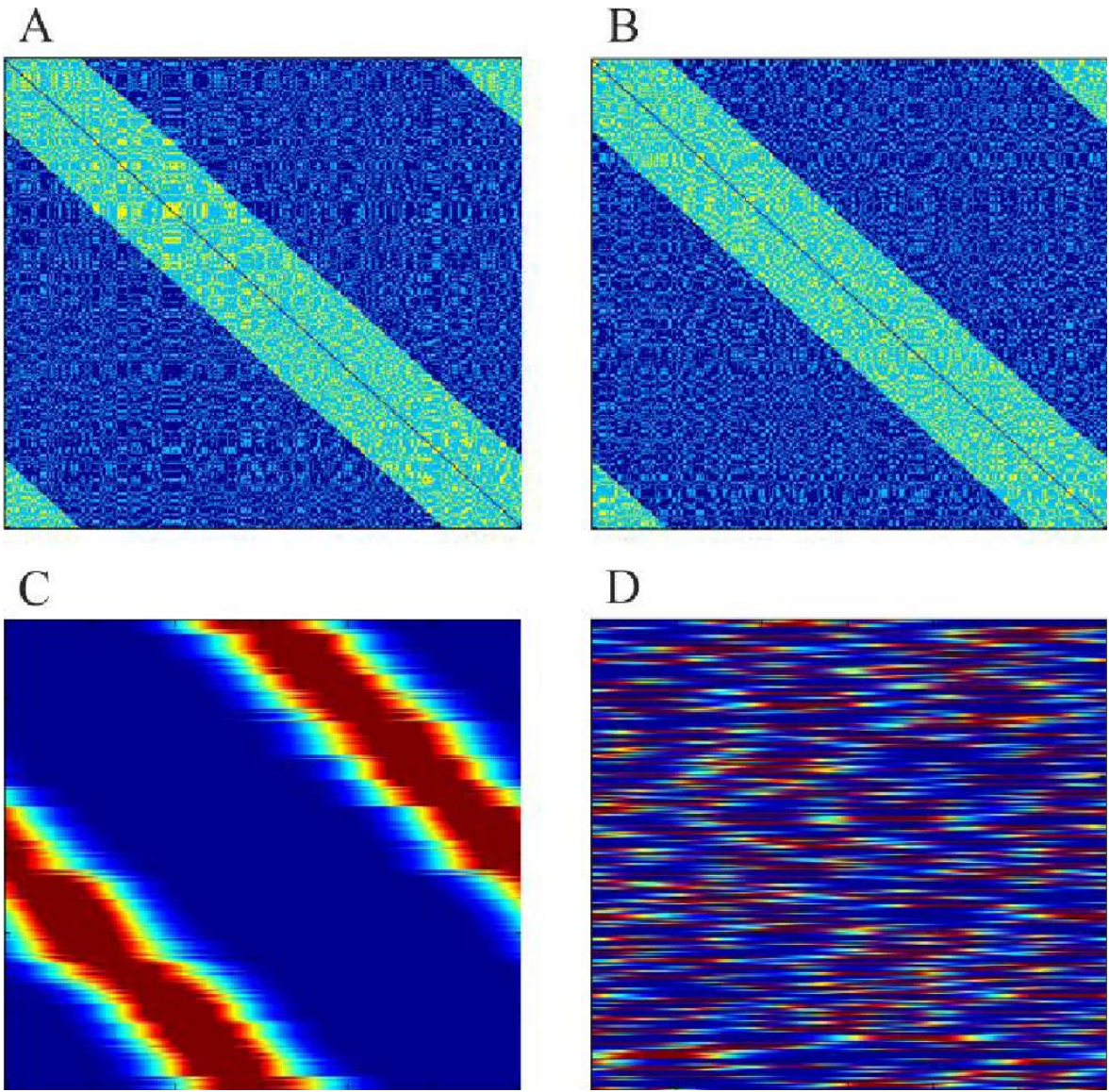

**Fig. 10.Neural network interconnection matrix T in the first (A) and in the second (B) enumeration**

The connection weight is color-coded: dark blue -10, light blue, 1, yellow 2. (C, D) the connection matrix W after completion of learning, (C) – neurons are numerated in the first enumeration; (D) – neurons are numerated in the second enumeration. The number of neurons $N$=300, the number of receptors $R$=300.

### 3.4.1 Leaky Integrate and Fire model

In this section, we compare the computed behavior of the neural networks of McCulloch-Pitts neurons described in previous sections with the behavior of the networks of Leaky Integrate-and Fire (LIF) impulse neurons .

Fig. 11 shows the activity dynamics in the network of LIF neurons. The excitatory neural network connections are made with the help of molecular markers, similar to the technique used with MCP neurons whose activity is shown in Fig. 7. The inhibitory neurons get excitatory connections from all excitatory neurons, and they send their connections back to all excitatory neurons. In Fig. 11, the color code of the pixels indicates the sum of the membrane potentials of the neurons in each of the $M$ standard neuron sets (SNS) (def. below). The horizontal coordinate is the order number of the SNS, while the vertical coordinate indicates time. The standard neuron sets are those, which include the neurons with the markers $(j - L/2)$, $(j - L/2 + 1), ..., j, (j + 1), ..., (j + L/2)$, for $j = 1, ..., M$, where $L = \delta$. The neurons have an accommodation property. Due to modeled calcium currents, the threshold of the neurons depends on their recent activity. Fig. 11 clearly shows that the activity of the network slides with time over the ring of the SNS.

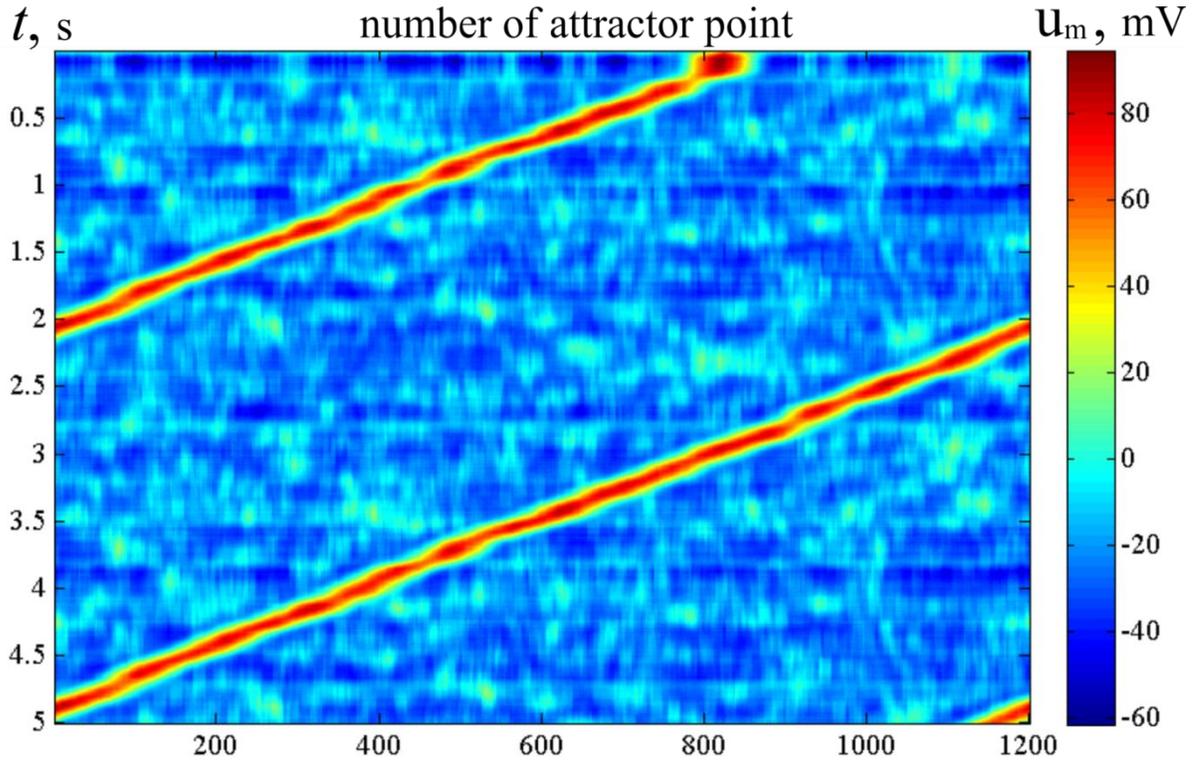

**Fig. 11. Raster of activity for LIF bump attractor**

Horizontal line - order number of the molecular marker. Vertical – time (s). The color of the dots codes the sum of membrane potentials of the neurons, which contains markers from $[x - L/2]$, to $[x + L/2]$. LIF neural network with accommodation (Supplementary Material 1.3), $\tau_{Ca} = 0.1$ c, $N = 600$, $M = 1200$, $L = 30$.

### 3.4.2 Dynamical neural attractors

In sections #2.2.1 (pre-formed bump attractors) and #3.4 we studied the structure of neural attractors with help of auxiliary dynamics, adding accommodation to neural properties. The dynamical properties of the neural networks can be set by appropriate learning [9 - 11] or forming of the connections [22]. Such networks can be also formed with the help of molecular markers if connection rules are made asymmetric. In simulations, connection rules in the network were analogous to the connection rules in #2.2.1. However, connected with excitatory connections were made only from neurons with larger order numbers of markers to the neurons with smaller order numbers, up to number difference of $L/2$ Fig. 12 demonstrates activity of such a type of neural network. The period of the activity in this case depends on threshold control and can be varied in some range (the early example of such a control is given in Fig. 1 of [10]).

It should be emphasized that the pattern of the activity of the neural network in Fig. 12 gives a clear picture of propagation of a wave in excitable media [33]. In Fig. 12 B one can see that the excitation wave form is asymmetric. Note that the upper-left edge of the excited area at Fig. 12 B (the front of excitation wave) looks more straight and sharp than the lower-right edge (the tail of the wave), alike the real wave. However, in this case, physically, there is no excitable media. The wave propagates, in fact, in the configurational space of the neural network. The existence of such kinds of waves was also observed in [9].

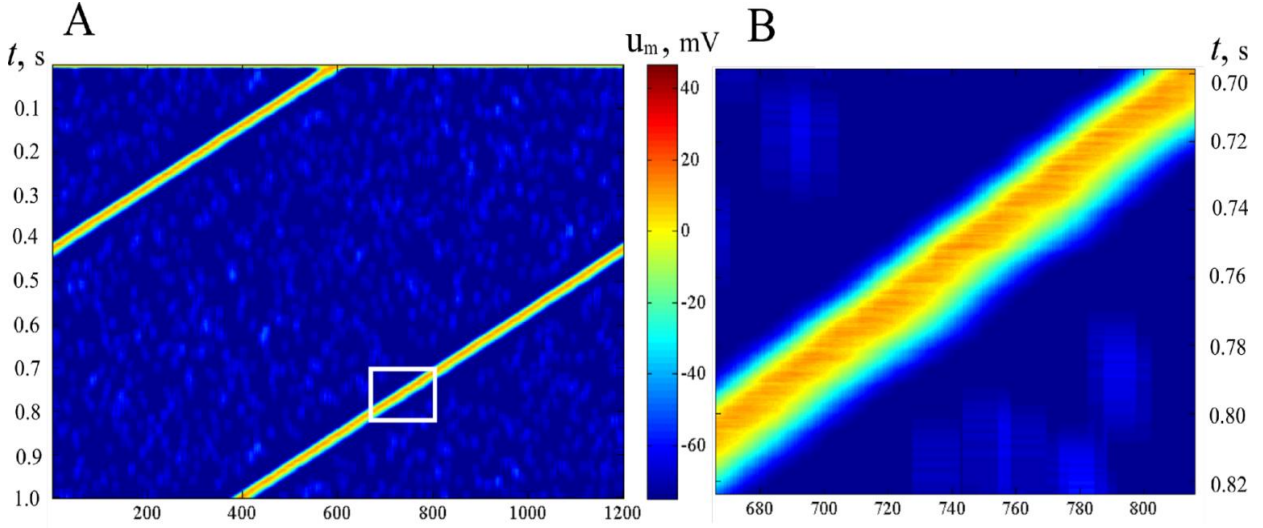

*Fig. 12. Activity propagation in LIF dynamic neural network*

Notations as in Fig. 11. (B) presents the rectangular inlet of (A). Asymmetric excitatory connections between neurons: from neurons with larger marker order numbers backwards. $N$=600, $M$=1200, $L$=30.

## 4. Discussions

### 4.1 Dimensions

From section 3.2. it is worth to make a note on properties of attractor networks with $d > 1$. In particular, it can be argued that the dimensionality of a reliable attractor neural network with a good resolution is hardly possible for $d > 4$; in any case, it can be shown that it cannot exceed the value $d = 8$.

Let us have a preformed continuous bump attractor of dimensionality $d$ with $d$-cube grid, constructed with the help of $d$-dimensional analogs of the molecular-marker method (see section #2.2.1). In this case, each neuron gets $k$ markers. With the help of each marker, the neuron is connected with $(2\delta)^d$ neurons (the volume of the $d$-dimensional neuronal neighborhood). So, each neuron has $(2\delta)^d \cdot k$ connections. This number must be less than the number of neurons: $(2\delta)^d \cdot k < N$. Let $l$ be a number of discrete elements for each dimension. Each element should be presented as a separate attractor state of the network. So, we have $l^d \leq N^2 / L$, or $d \leq \log(N^2 / L) / \log(l)$. The estimate of maximal value of $d$ can be obtained in the following way. First, note that this estimate at fixed $N$ increases with the decrease of $L$ and $l$. For the number of neurons in the range $N = 10000 \div 50000$ (the usually supposed number of neurons in one cortical column in human brain), we get $d = 7 \div 8$ for $l = 10, L = 10$ and $d = 3 \div 4$ for $l = 100, L = 100$. In detail we have studied here only the cases of $d$=0 and $d$=1.

### 4.2. Two layers perceptron implemented with isolated points attractor

In sections 2.1 and 3.1 we have introduced and explored neural networks with inborn point attractors ($d$=0), Properties of these attractor neural networks are similar to properties of Hopfield networks. We have revealed the functionally important properties of these attractors. When we have constructed a perceptron, the neural network, which detects specific patterns, it occurred that attractor-based perceptrons substantially surpass neuron-based perceptron's error tolerance (see Fig. 3 B).

## 4.3 Molecular Marker Based  Neural Network with 1-d bump attractor

In this paper we have developed a physiologically possible mechanism to make connections in the network to have in it a "long" one-dimensional attractor, i.e. the attractor with the number of states, exceeding the number of neurons in the network. We have demonstrated examples of such networks, obtained in computational experiments. Our examples can serve as templates for interpretation of neurophysiological experiments, as one-dimensional neuron attractors are often supposed to be present in different brain structures. For the experimental verification purposes we have proposed two methods for visualization of neural attractor activity. The methods were tested on data of computational experiments and might be used for physiological experiments.

## 4.4 Extension of Kohonen's SOM

We explored the bump attractor-based SOM,  similar to Kohonen's original construction (preliminary results were published earlier [30]). The construction  demonstrates a series of considerable distinctions from the original Kohonen's paradigm. First of all, it is more likely to be implemented in a real brain than the original construction of Kohonen, as Kohonen's chains and grids [31] have no analogs in real systems.  Second, the neural activity regenerative processes in bump attractor network, provides the natural neighborhood of the attractor states by the mere nature of the bump attractor. Indeed the neighborhood of a given state consists of the states, which are close to a given state by Hamming metrics. In the Kohonen's SOM, the neighborhoods of the nodes are defined forcefully.

Thus, computational experiments on the learning of neural networks with bump attractors to respond to external signals have demonstrated that SOM-like mechanisms can be efficient for representation of continuous variables in realistic neuronal systems.

## 4.5 Neural Building Blocks

It was believed for a long time that neural information processing and neural control is based on a set of "principles of neural organization", just as the artificial information and control machines use standard operations and standard circuits [6, 37 - 40]. There is hope that the number of such concrete circuit principles is, although large, is not huge (say, less than 1000). One of the ways to complete the "brain reverse engineering" [19] is to reveal these principles, one by one, in order to obtain a complete set of them [18]. The bump attractor neural structures (with $d$=0 and $d$=1) and their versions definitely constitute a part of this set of principles [34]. In this paper, we have presented only fragments of the future detailed description of the ways of functioning and functions of bump attractors. There is still a long way to go until discovery of new principles and for this knowledge to be applied in future artificial mind systems.

## 4.6 The Clock-Brain Machinery

We hope that hereby we have demonstrated that the simple operation of switching on of stable static or dynamic recurrent states, provided by inborn neuronal connections, can serve as a basic computational operation of the neural systems in many cases. In this capacity, the equal utility might be assumed for the neural networks with attractor dimensions $d = 0, 1, ..., \leq 8$. However, it is possible that the case of $d = 1$ is in fact special and might be considered a basis for a large class of interdependent constructions which enable the effective functioning of the neuronal systems. On one hand, such a conclusion might be based on the fact that any type of multi-dimensionality can be considered as a Cartesian product of the appropriate number of one-dimensional constructions. On the other hand, there emerges a new class of structures in neuronal systems that is based on the networks with $d = 1$. We would propose to name this class "The Clock-Brain Machinery". What does this mean? It would be relevant to recollect the remark of Masao Ito that states that the purpose of understanding concrete human-made mechanisms in the long history of their existence in human culture has been, in essence, for disassembling and further successful reassembling of these mechanisms. For centuries, the most sophisticated of them were the mechanical clocks [40]. Ito expresses the opinion that the likely considerations should be used for analysis of the brain machinery. This note might be treated as allegory, but there are arguably more concrete elements in it as well. We suppose that the dynamic ring attractor neural network (Fig. 12) can serve as a basic block of the realistic clock metaphor. One should also bear in mind that static attractor one-dimensional structures can be easily transformed into dynamic one-dimensional attractor structures [11, 24] with switching on of the Ca++-dependent potassium channels. The complete cycle of activity in the network of the type of Fig. 12 resembles the full turn of a clock's gear wheel. The activity of different cyclic networks can be easily connected, yielding a structure resembling the gear-wheeled clock mechanism. Of course, it should be taken into consideration that specific states of neural "gear wheels" can have different meanings. Of course, the clock-wise picture of the brain machinery doesn't pretend to present all of a real brain's processes, but it might yield a convenient framework for getting more details of the brain function.

## 5. Conclusion

In this paper we extend our work on attractor based neural networks. Special attention is paid here to aspects of neural dynamics insufficiently highlighted beforehand. In this paper, we have demonstrated that robustly functioning neural networks can have $M = k \cdot N$ attractor states, where $k = 1 \div 1000$. On the other hand, there is a possibility, that the value of k as low as $10^{-2}$ - $10^{-4}$. For example, it is believed that the typical "grandma neuron" is definitely a representative of a sub-network, containing $N_{grandma} = 10^2$ - $10^4$ neurons [35]. The latter assumption cannot be overturned. If the neurons, which take part in one "grandma" representation are not used for other purposes, then we get the very low estimate of $k = 1 / N_{grandma}$. As a result, one of the main problems in experimental and theoretical neural studies is to understand if neuronal elements in brain are used to represent only unique external events, or if they can be used in combinations, so that different combinations of active neurons can represent different external events. The fact of the existence of remapping in the hippocampus [36] shows that at least in some neural structures, the combinations are used. For the attractor neural networks, we have considered in a general form the networks whose set of attractor states might represent finite grids for discrete and continuous variables. This treating naturally leads us to notions of neural networks with attractor dimensions $d = 0, 1, 2$, etc. We give simple calculations, showing that for biological neural networks $d$ is (fussy) limited, $d = \leq \sim 8$.

Further, we consider a method ("molecular markers") for forming inborn connections in neural networks, which provide neural networks with attractor dimensions $d=0$ and $d=1$. The $k$ value in the cases we have considered is in the range $k = 1 \div 1000$. The elaboration of inborn neural networks with attractor dimensions $d>1$ is an interesting problem for future works. Also, we note that the activity of neural networks with inborn attractors can obtain meaning due to inborn or learned forming of connections of a concrete neural network with other neural structures. In pursuing this idea, we showed our results for the cases of $d=0$ (two-layer perceptrons) and $d=1$ (Kohonen's SOM, based on one-dimensional attractor states). The last part of our paper demonstrates that the results, obtained with McCulloch-Pitts neural model, have direct analogies to the networks of impulse neurons of the LIF type. There is no reason to assume that it doesn't hold true for more detailed physiological neural models. This model may find applications in many areas including deep neural nets for image, text and voice recognition, autonomous driving and drug discovery and drug repurposing and provide a theoretical base for further research in applying biological principles to machine learning.

**Supplementary Materials:**

1. The neural models
2. Number of inborn attractor states for d=0
3. Distances between states in 1 d bump attractor obtained with help of model molecular markers
4. Evaluation of kc in one-dimensional bump attractors
5. Trade-off relations for network attractors

**Acknowledgements:**

Authors would like to thank Abhaya Chandra Kammara Subramanyam at ISE, TU Kaiserslautern, Kaiserslautern, Germany for valuable contributions and edits to the manuscript. Also, the authors are thankful to V.B. Kotov for many useful comments to the manuscript. The work was supported in part by RFBR grant # 13-07-01004 to WLDB

## Supplementary Material 1. The neural models

We use several types of neural models in calculations. Hereby the main features of the models are described in detail.

### A1.1. McCulloch-Pitts (MCP) neural model.

In this case, the behavior of the $i$-th $(i = 1, ..., N)$ neuron is described by its phase function, $\varphi_i(t)$.

There are two constants: the duration of excitation, $w$, and the duration of refractoriness, $r$. If $\varphi_i(t) \geq r$ and $(\sum_{j=1}^{N}(w_{ji} \cdot x_j(t)) - \theta_i) \geq 0$, then $\varphi_i(t+1) = 0$, otherwise $\varphi_i(t+1) = \varphi_i(t) + 1$. The output of the neuron $x_i(t) = 1$, if $\varphi_i(t) < w$, otherwise $x_i(t) = 0$. Usually, $w = 1$ and $r = 0$.

The important rule for the MCP neural networks is the rule of updating the values of $\phi_i(t)$ and $x_i(t)$. The rule, described above, is known as synchronous dynamics: the phases of all neurons are updated simultaneously. Sometimes, we used the asynchronous random dynamics. In this case, the updating is performed in cycles of $N$ updates. In each cycle of updates, the order of neurons is selected randomly and, in this order, neurons are updated one by one; the freshly updated neuron takes part in updating the next neurons in that cycle of $N$ updates.

A1.2. LIF model
The dynamics of these neurons is described by the following equation

$$\tau_m \frac{du_m(t)}{dt} = -\left(u_m(t) - u_r\right) + R_m \left(i_{syn}(t) - i_{inh}(t) - i_{Ca}(t)\right) \qquad \text{(A1.1)}$$

with

$$\tau_{Ca} \frac{di_{Ca}(t)}{dt} = -i_{Ca}(t), \qquad \tau_{Ca} = 0.1 \text{ s} \qquad \text{(A1.2)}$$

Here, $u_m(t)$ - membrane potential, $\tau_m = R_m C_m$ - membrane time constant, $C_m = 1 \text{ nF}$ and $R_m = 10 \text{ M}\Omega$ - capacitance and resistance of the membrane, $u_r = 0 \text{ mV}$ - resting potential of the membrane, $i_{syn}(t)$ - the sum of excitatory synaptic currents, $i_{inh}(t)$ - the inhibitory current, $i_{Ca}(t)$ - the accommodation current, which depends on activity of neurons. This current is considered to reflect the action of Ca$^{++}$-dependent K$^+$ channels. With each spike $i_{Ca}(t)$ gets an increase of 1 nA. The neuron generates an impulse (spike) when its membrane potential $u_m(t)$ attains threshold $u_{th}$. Afterwards, its membrane potential momentarily jumps to zero value, while threshold gets increase of 1000 mV and then decreases according to equation:

$$\tau_{th} \frac{du_{th}(t)}{dt} = -\left(u_{th}(t) - u_{th0}\right) \qquad \text{(A1.3)}$$

with $\tau_{th} = 2 \text{ ms}$, and $u_{th0} = 10 \text{ mV}$. Each incoming impulse momentarily adds 0.3 nA to $i_{syn}(t)$. In the absence of incoming impulses, $i_{syn}(t)$ decays exponentially to zero with time constant of 25 ms.

The inhibitory current, $i_{inh}(t)$, is the same for all excitatory neurons in the network and represents a global variable which controls the activity of the network. It is controlled by the following equations:

$$\tau_e \frac{di_e(t)}{dt} = -i_e(t) + N \sum_j \sum_f \delta(t - t_j^f) \qquad \text{(A1.4)}$$

$$\begin{cases} i_{inh} = k_{inh} \left(i_e(t) - I_{e0}\right), & if \quad i_e(t) > I_{e0}, \\ i_{inh} = 0 & otherwise \end{cases} \qquad \text{(A1.5)}$$

Variable $i_e(t)$ represents the excitatory input to inhibitory neurons, $\tau_e = 10 \text{ ms}$, $k_{inh} = const$, $I_{e0} = 10 \text{ nA}$. The synaptic delays between neurons were uniformly randomly distributed in the range $1.0 - 5.0$ ms.

**Supplementary Material 2. Number of inborn attractor states for d=0**
The mean value of matrix elements is:

$$\gamma = \overline{T_{ij}} = \Pr\{T_{ij} = 1\} = 1 - \Pr\{T_{ij} = 0\} = 1 - \left(1 - \frac{L^2}{N^2}\right)^M. \qquad \text{(A2.1)}$$

We suppose that $M$ is proportional to $N^2 / L^2$, i.e. $M = \kappa \cdot (N^2 / L^2)$. When $M \to \infty$, we have:

$$\gamma = 1 - \left(1 - \frac{\kappa}{M}\right)^M = 1 - e^{-\kappa}. \qquad \text{(A2.2)}$$

Now, let the network have at the input one of its theoretical attractor patterns. Then, the probability, that a 'foreign' neuron have $L$ excitatory inputs is

$$P\{h_i = L\} = \gamma^L.$$

So, the probability that at least one of $(N$-$L)$ foreign neurons will get $L$ units of excitation is:

$$P_{err} = P\{\exists i_{inactive} : h_i = L\} = 1 - (1 - \gamma^L)^{N-L} \approx 1 - (1 - \gamma^L)^N$$

Taking $P_{err} = 1/N$, we have:

$$1 - (1 - \gamma^L)^N = \frac{1}{N},$$

and, after non-complicated transformations, leaving only first order (by $1/N$) terms, we have:

$$\gamma = N^{-2/L}. \qquad (A2.3)$$

Comparison of (A2.2) and (A2.3) finally yields:

$$\kappa = -\ln\left(1 - N^{-2/L}\right). \qquad (A2.4)$$

Table A2 gives numerical values of $\kappa$ for a set of $N$ values at $L$=20. Although computational estimates of $\kappa$ give the constant value $\kappa = 1$, one can see that the analytical reasoning does not diverge too far from the computational experiments estimate. Thus, Table A2 shows that theoretical estimates of $\kappa$ for $N \leq 5000$ are less than two times smaller than 1.0.

| $N$ | 100 | 500 | 1000 | 2000 | 3000 | 500 |
|---|---|---|---|---|---|---|
| $\kappa$ | 1.0 | 0.77 | 0.69 | 0.63 | 0.6 | 0.56 |

## Supplementary Material 3. Distances between states in 1 $d$ bump attractor obtained with help of model molecular markers

Fig. A1 presents a fragment of the layout of all neurons in order of the order numbers of the markers, which they contain (as each neuron has k markers, each neuron is presented $k$ times in this layout). The neurons become excited in the same order, while the activity propagates over the bump attractor. We now consider the distance between some initial state, $X_0$ and states which are following it, $X_1, X_2, X_3, \ldots X_t$, ($t$=1, 2, … M).

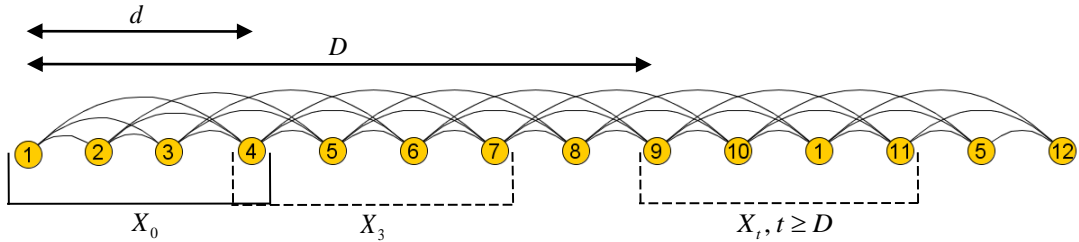

**Fig. A1**

Fig. A2 shows the inner product $\langle X_0, X_t \rangle$ of two states plotted against t. It is obvious that in the beginning, the intersection between $X_0$ and $X_t$ decreases linearly from $L$ to 0. Then, it stays 0 up to $t = D$. For $t > D$, the intersection is random. Its mean value can be obtained as follows. Each neuron of the first state takes later part in ($k$-1) states That means that for any neuron which is excited in $X_0$, the probability to be excited in $X_t$ (for large $t$) is

$$p = \frac{(k-1)L}{M - L}.$$

That means that the average intersection between $X_0$ and $X_t$ will be $pL \approx (k-1)L^2/M$

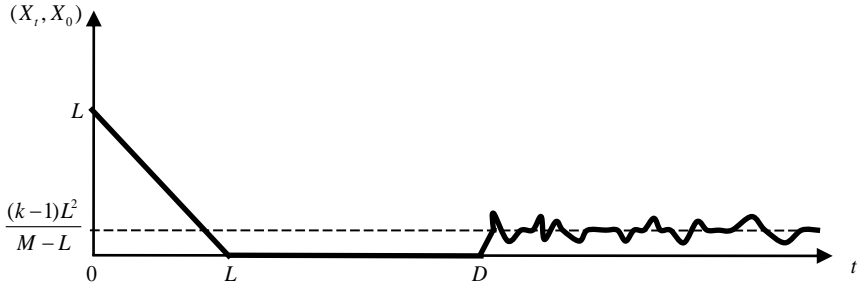

**Fig. A2**

The Hamming distance between the states is connected with intersection of them according to the relation $H(X_t, X_0) = 2(L - (X_t, X_0))$, where from we get for $t \geq D$

$$\underline{D} \approx 2L\left(1 - \frac{(k-1)L}{M}\right).$$

With $L=15$, $k=3$, $M=900$, we have $\underline{D} \approx 29$, which coincides with the results of the computational experiments.

## Supplementary Material 4. Evaluation of $k_c$ in one-dimensional bump attractors

As each neuron takes part in $k$ attractor states, each line of the matrix $T$ contains about $2\delta k$ positive (equal to 1) matrix elements. The probability of positive matrix element is $2\delta k / N$. That means that the probability of firing of one excessive neuron, which is not active in $X_0$, is $(2\delta k / N)^L$. On the contrary, the probability that this will not happen for any of the remaining $N - L$ neurons is $\left[1 - (2\delta k / N)^L\right]^{N-L}$. In computational experiments, the critical value $k_c$ was defined as the value of $k$, such that five random networks with a given $k$ show perfect cycles. Thus, for $k_c$ we obtain an equation:

$$1 - \left[1 - \left(\frac{2\delta k_c}{N}\right)^L\right]^{N-L} < \frac{1}{5}$$

And finally, for the critical value of $k$, we have:

$$k_c < \frac{N}{2\delta \sqrt[L]{5N}}, \qquad (A4.1)$$

Factor $\sqrt[L]{5N}$ is of the order of 1 and changes very slowly with $N$. So, (A4.1) yields practically linear dependence of $k_c$ on $N$. (cf. Fig. 17). In particular, we have:

at $L = 15$, $N = 300$, $\sqrt[15]{5 \cdot 300} \approx 1.62$, k = 7.7 (in experiment, k = 5);

at $L = 15$, $N = 1200$, $\sqrt[15]{5 \cdot 1200} \approx 1.79$, k = 28 (in experiment, k = 21).

## Supplementary Material 5 Trade-off relations for network attractors

This section uses the computational experiments first presented to the conference "Neuroinformatics 2013" [25]. For neural network applications, it is important to know how the network behaves depending on the values of its parameters. In particular, it is important to know how stable the attractor points are. Fig. A3 shows the "error" in states, as a function of $N$, $M$ at $L = 20$, obtained in Monte Carlo computational experiments. The synchronous $L$ winners dynamics (Supplementary Material 1) of the network was used and the "error" was considered to be the Hamming distance between the experimentally obtained stable state and the "theoretical" attractor point which served as the initial condition. It can be seen that the error grows with $M$. For $M$ less than a critical value, $M_{cr}$, the "theoretical" attractor points are stable.

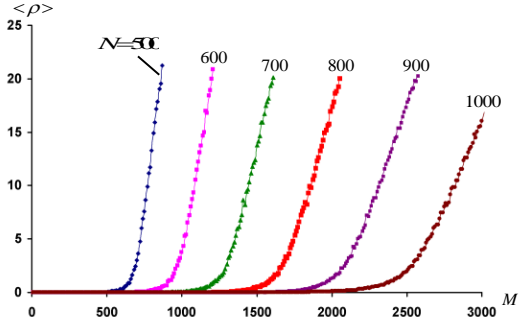

**Fig. A3. The "error" in neural network stable states**

N and M are the number of neurons and the number of marker types. Each type of markers has the same number, $L$=20, of elements. In each case, plotted are values averaged over all $M$ "theoretical" attractor points.

Fig. A4 shows dependence of $\alpha_{cr} = M_{cr} / N$ on $N$. The linear empirical approximation shows that with increasing N, the value of $M_{cr}$ increases $\approx N^2$ (when $L$ is constant): $M_{cr} \approx (N / L)^2$.

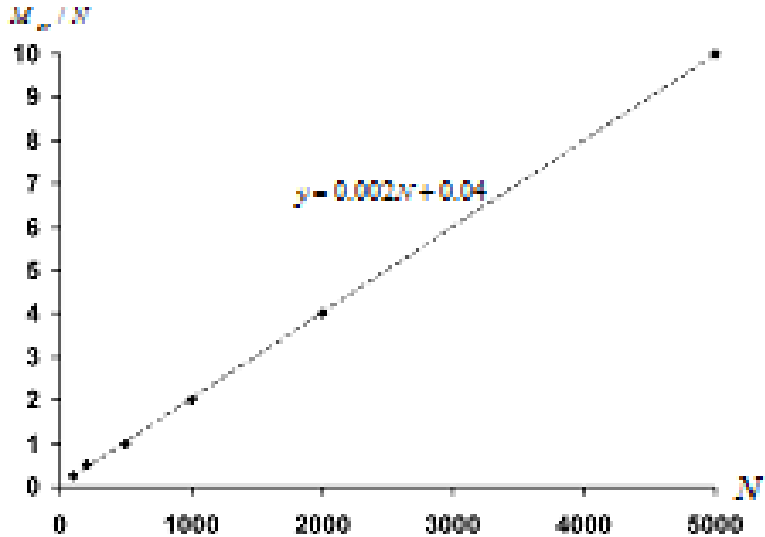

**Fig. A4. The critical values of ratio $M_{cr} / N$ as a function of $N$**

Computational experiments (dots); broken line is the least square regression. $L$=20.

The mean value of matrix elements is

$$\overline{T_{ij}} \cong ML^2 / N^2 \qquad (1)$$

When network resides at attractor state, $S^m$, the $L$ neurons which are active get the following inputs:

$$h_{act} = \sum_{j=1, j \neq i}^{N} \mathbf{T}_{ij} S_j^{\,m} = L - 1 \qquad (2)$$

Input to the rest ($N - L$) of the neurons is approximately:

$$h_{inact} = L \cdot \overline{\mathbf{T}_{ij}} \qquad (3)$$

Here $\overline{\mathbf{T}_{ij}}$ is the average value of matrix element of $\mathbf{T}$. The distinction between the right parts of (2) and (3) enables the neural network discriminate between attractor and non-attractor states. Supplementary Material 2 gives the analytical reasoning, which qualitatively explain the data of computational experiments, displayed in Figs. A3 and A4.

It is known that not all sets of points can be separated by a plane into two subsets. For example, any four points in 3-d space can be divided by a plane in any combinations, but for five points that can not always be done.

A general result related to this problem has been obtained by E. Gardner [27]. It states that if there are randomly chosen $2R$ points in the R-dimensional space, painted randomly in two colors, $R$ points in each group, then at $R \to \infty$, with probability approaching 1, exists the plane which separates the colored points by colors. But when the number of points exceeds $2R$, the probability of separation approaches 0 as $R \to \infty$. This result states that at maximum, we can separate only $2R$ points in $R$-dimensional space. In the case of separating a few points from the others, the situation changes. Let us have M points in R-dimensional space. Let then divide them into two parts: k points in one of them and $M - k$ in the other. In this case (formula (40) of [27]):

$$\alpha_c = -\frac{1}{(1-m) \cdot \ln(1-m)} \ .$$

Where, in our case, $\alpha_c$ and $m$ are correspondingly $\alpha_c = M / R$ and $m = 1 - 2k / M$. So the expression for relations between $R$, $M$, and $k$ is:

$$R = 2k \cdot \ln(M / 2k) \qquad (4)$$

We tested this theoretical estimation in computational experiments for a set of different values of $k$ and $R$ (Fig. A5):

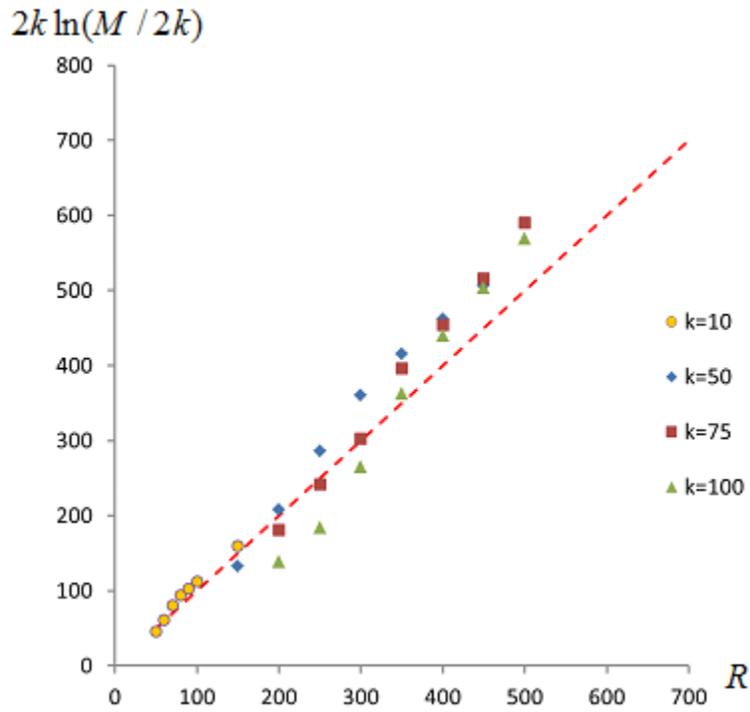

**Fig. A5. Data of computational experiment on relation between *M*, *R* and *k***

The linear separation was performed with the help of Rosenblatt algorithm. The broken line corresponds to equation (4). Parameters are shown in the figure.

So, the number of patterns *M* when *k* is fixed is an exponent of *R*. Further, we use $k = (ML/N) \approx 0.01M$, and *R* in the range 100 – 500.